\ifdefined\WORDCOUNT
\documentclass[superscriptaddress,twocolumn,preprintnumbers,amsmath,amssymb,prl,nofootinbib]{revtex4-1}
\else
\documentclass[superscriptaddress,twocolumn,showpacs,preprintnumbers,amsmath,amssymb,prl]{revtex4-1}
\fi
\usepackage[table]{xcolor}

\newcolumntype{s}{>{\columncolor[HTML]{AAACED}} p{3cm}}

\usepackage{amssymb}
\usepackage{epsfig,amsmath,graphics}
\usepackage{epstopdf}
\usepackage{graphicx}
\usepackage{color}
\usepackage{gensymb}

\usepackage[utf8x]{inputenc}

\usepackage{natbib}
\usepackage{graphicx}

\begin{document}

\title{A 40 W, 780 nm laser system with compensated dual beam splitters for atom interferometry
}

\author{Minjeong Kim}
\affiliation{Stanford University, Stanford, CA 94305}
\author{Remy Notermans}
\altaffiliation[Current address: ]{Atom Computing, Inc., Berkeley, CA 94710}
\affiliation{Stanford University, Stanford, CA 94305}
\author{Chris Overstreet}
\affiliation{Stanford University, Stanford, CA 94305}
\author{Joseph Curti}
\affiliation{Stanford University, Stanford, CA 94305}
\author{Peter Asenbaum}
\affiliation{Stanford University, Stanford, CA 94305}
\author{Mark A. Kasevich}
\email{kasevich@stanford.edu}
\affiliation{Stanford University, Stanford, CA 94305}
\date{August 3, 2020}

\begin{abstract}
We demonstrate a narrow-linewidth $780\; \text{nm}$ laser system with up to $40\; \text{W}$ power and a frequency modulation bandwidth of $230\; \text{MHz}$. Efficient overlap on nonlinear optical elements combines two pairs of phase-locked frequency components into a single beam. Serrodyne modulation with a high-quality sawtooth waveform is used to perform frequency shifts with $> 96.5 \%$ efficiency over tens of MHz. 
This system enables next-generation atom interferometry by delivering simultaneous, Stark-shift-compensated dual beam splitters while minimizing spontaneous emission. 
\end{abstract}

\maketitle

Narrow-linewidth, high-power laser systems are of substantial interest due to their applications in atomic physics and precision measurement;  in particular, the performance of precise atom interferometers is often limited by the characteristics of the laser that diffracts and interferes the atoms \cite{parker2018measurement, asenbaum2020atom, stuhler2003magia, graham2013new}.  Such experiments require high-power lasers with low phase noise \cite{asenbaum2017phase}, frequency agility of tens to hundreds of MHz \cite{overstreet2018effective}, and an optical spectrum that contains frequency components separated by tens of GHz or more \cite{kovachy2015quantum}.
Low-noise, frequency agile lasers with more than $1.6\; \text{W}$ of power \cite{muller2006phase} and compact laser systems \cite{merlet2014simple, fang2018realization} have been demonstrated for such applications, each fulfilling some of the necessary requirements. Other useful approaches include frequency shifting using serrodyne modulation \cite{cumming1957the, johnson2010broadband} and generation of tens of Watts at 780 nm \cite{Sane2012, chiow2012generation}. However, it is challenging to implement all of the criteria discussed above in a single laser system.

For realization of an atom interferometer with Stark shift compensated dual beam splitters, it is necessary to have four different frequency components:  two for the red-detuned Bragg pair, and two for the blue-detuned Bragg pair \cite{kovachy2015quantum}.  The Bragg pairs should have equal and opposite detunings from the single-photon resonance so that the overall Stark shift nearly vanishes, and the frequency difference within each pair must be swept over tens of MHz to meet the two-photon resonance condition for the Doppler-shifted Bragg transition on each interferometer arm. This scheme should also work for Raman beamsplitters.

\begin{figure}[!b]
    \centering
    \includegraphics[width=\linewidth]{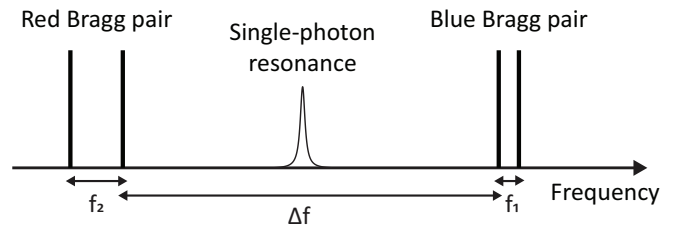}
    \caption{Optical spectrum. The spectrum consists of two pairs of frequencies (Red Bragg pair and Blue Bragg pair) on either side of the single-photon resonance. In the configuration described in this work, $\Delta f = 370\; \text{GHz}$.  The frequency differences $f_1$ and $f_2$ can be tuned independently by tens of MHz. Figure is not to scale.
    }
    \label{fig:spectrum}
\end{figure}

In this Letter, we demonstrate a $40\; \text{W}$, $780\; \text{nm}$ laser system with high frequency agility and low phase noise.  The $780\; \text{nm}$ light is produced by frequency doubling amplified $1560\; \text{nm}$ lasers.  We efficiently overlap four different frequency components during and after the doubling process to generate an optical spectrum that spans $370\; \text{GHz}$.
The laser wavelengths are centered around $780.24\; \text{nm}$, which is the rubidium D$_2$ resonance, and the frequency span can be varied by up to $2\; \text{GHz}$. 
The frequency of each component can be shifted by serrodyne modulation of the 1560 nm laser with efficiency $> 96.5\%$ over tens of MHz.  Phase stabilization of the $780\; \text{nm}$ light is accomplished by phase shifting of the $1560\; \text{nm}$ lasers.
Fig.~\ref{fig:spectrum} shows a schematic of the  optical spectrum generated by the laser system, which meets all of the requirements for implementing Stark shift compensated dual beam splitters.

\begin{figure*}[t]
\centering
\includegraphics[width=\linewidth]{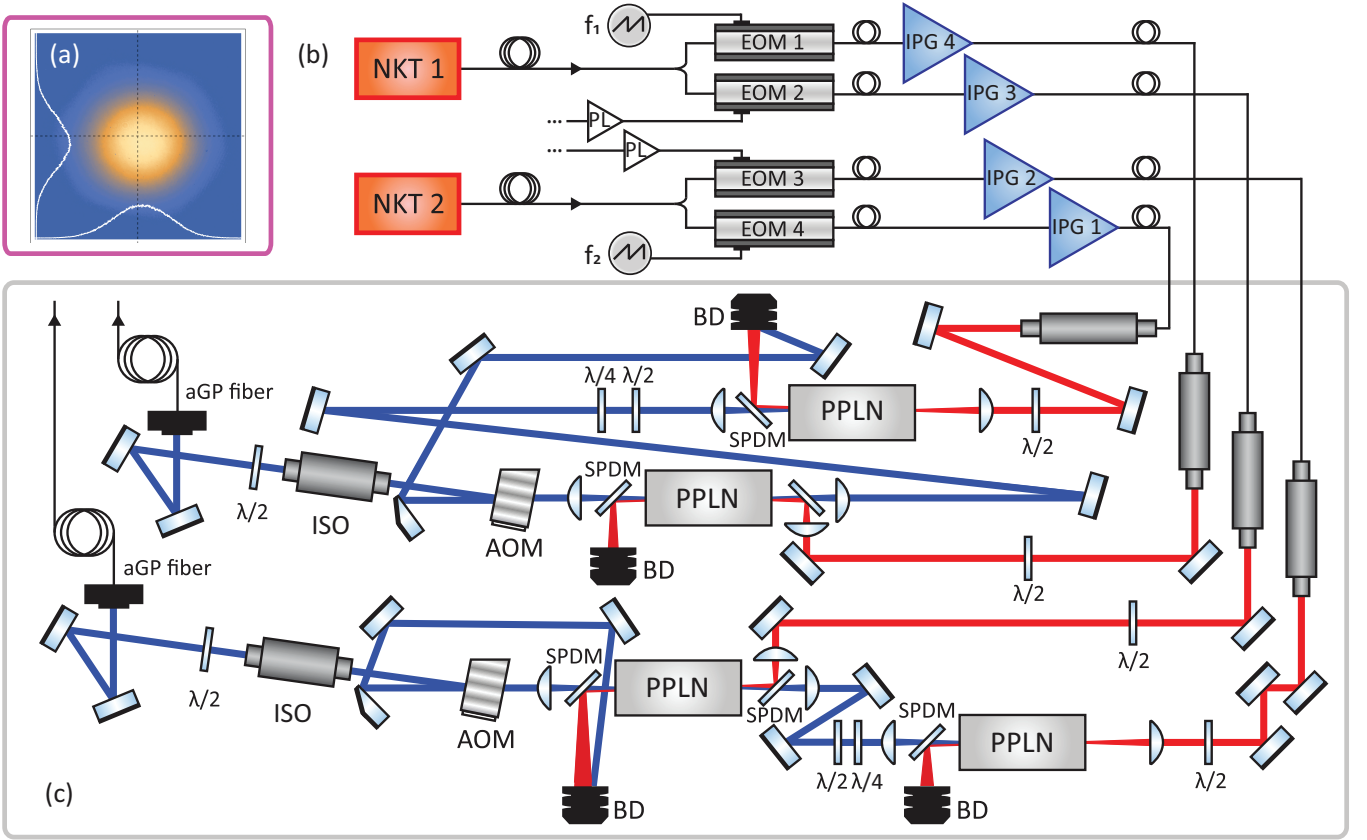}
\caption{Schematic of the laser system. The (red, blue) beams denote ($1560\; \text{nm}$, $780\; \text{nm}$) light. Figure is not to scale. NKT: NKT Photonics fiber laser, EOM: electro-optic modulator, PL: phase lock, IPG: IPG Photonics fiber amplifier, $\lambda$/2: half-wave plate, $\lambda$/4: quarter-wave plate, PPLN: periodically-poled lithium niobate crystal, SPDM: short-pass dichroic mirror, BD: beam dump, AOM: acousto-optic modulator, ISO: Faraday isolator, aGP fiber: aeroGuide-Power polarization-maintaining single-mode fiber. 
(a) Beam profile of $780\; \text{nm}$ light at $6\; \text{W}$ produced by IPG 1 at a distance of $100\; \text{cm}$ after the aGP fiber output collimator;
(b) serrodyne modulation setup;
(c) frequency doubling optics.
}
\label{fig:layout}
\end{figure*}

We use two $1560\; \text{nm}$ fiber lasers (NKT Photonics Koheras BASIK X15) as the master lasers (Fig.~\ref{fig:layout}).  Each fiber laser has a fixed output power of $30\; \text{mW}$ and a specified linewidth of less than $0.1\; \text{kHz}$. The two wavelengths are centered around $1560\; \text{nm}$ and are $1.5\; \text{nm}$ apart. Each laser output is split $50:50$ and coupled into a pair of electro-optic modulators (EOM, EOSpace PM-5S5-10-PFA-PFAP-UV) with half-wave voltages $\text{V}_{\pi}\lesssim 3\; \text{V}$ at $1\; \text{GHz}$. One of the two EOMs in each path is used for frequency shifting, and the other is used for phase locking. Each EOM output is sent to a fiber amplifier (IPG Photonics EAR-30-LP-SF) and is amplified to a maximum power of $30\; \text{W}$.

The outputs of fiber amplifiers seeded by different NKT lasers are paired and frequency doubled using the optics configuration shown in Fig.~\ref{fig:layout}.  Each beam in each pair is frequency doubled in a separate PPLN nonlinear crystal (Covesion custom MSHG1550-1.0-40) \cite{chiow2012generation}. 
The doubling efficiency is maximized by individually optimizing the temperature of each crystal. 
The crystals are mounted inside ovens (Covesion PV40) for thermal stability and are temperature-controlled using the Covesion OC2 controller.

The output of each PPLN crystal passes through a shortpass dichroic mirror (SPDM) to deflect the remaining $1560\; \text{nm}$ light into a beam dump. 
Each $780\; \text{nm}$ beam proceeds into a plano-convex lens and is collimated to a waist of $1.2\; \text{mm}$. 
A half- and a quarter-wave plate are used for polarization birefringence compensation.

To overlap the two $780\; \text{nm}$ frequency components on each path, the $780\; \text{nm}$ output from the first PPLN crystal is spatially overlapped on a SPDM with the $1560\; \text{nm}$ light that is directed into the second crystal.  Since the doubling bandwidth of the PPLN crystal is $\sim 30\; \text{GHz}$ and the $780\; \text{nm}$ light from the first crystal is detuned by $370\; \text{GHz}$, the $780\; \text{nm}$ light from the first crystal passes through the second crystal without substantial alteration as the $1560\; \text{nm}$ light is frequency doubled.  The output of the second crystal consists of a beam that contains both $780\; \text{nm}$ frequency components.  This method provides an efficient way to overlap the two $780\; \text{nm}$ frequencies without much loss. 
The total $780\; \text{nm}$ power in both paths after doubling at full power is up to 40 W.

On each beam path, an acousto-optic modulator (AOM) controls the beam intensity. The diffracted beam continues along the path while the undiffracted beam is picked off and directed to a beam dump. A RF synthesizer (Moglabs Agile RF Synthesizer RF-421) supplies the $80\; \text{MHz}$ RF signals that drive the two AOMs. The amplitude of each RF signal is shaped by a voltage-controlled attenuator (Mini Circuits Frequency Mixer ZP-1H+, ZP-1LH) that is controlled by an arbitrary function generator (Tektronix AFG3102). This technique allows the production of laser pulses with arbitrary shape. 

The diffracted beam from each AOM is sent through a Faraday isolator to protect from back reflections and then coupled into a $5\; \text{m}$ polarization-maintaining single-mode fiber (NKT Photonics aeroGuide-Power) with $\sim 75\%$ coupling efficiency. This fiber has a large mode field diameter ($12.5\; \mu \text{m}$) 
and customized collimators (OZ Optics, HPUCO-15-780-P). Fig.~\ref{fig:layout}(a) shows the beam profile of the fiber output at a distance of $100\; \text{cm}$ from the collimator. To image the $6\; \text{W}$ beam, the beam was pulsed on for $10\; \mu$s at a repetition rate of $1\; \text{Hz}$. The weak hexagonal structure in the tails gets less pronounced while propagating due to its higher divergence.  Otherwise, the beam profile is Gaussian and has no noticeable fringes or aberrations.

\begin{figure}[tb]
\centering
\includegraphics[width=\linewidth]{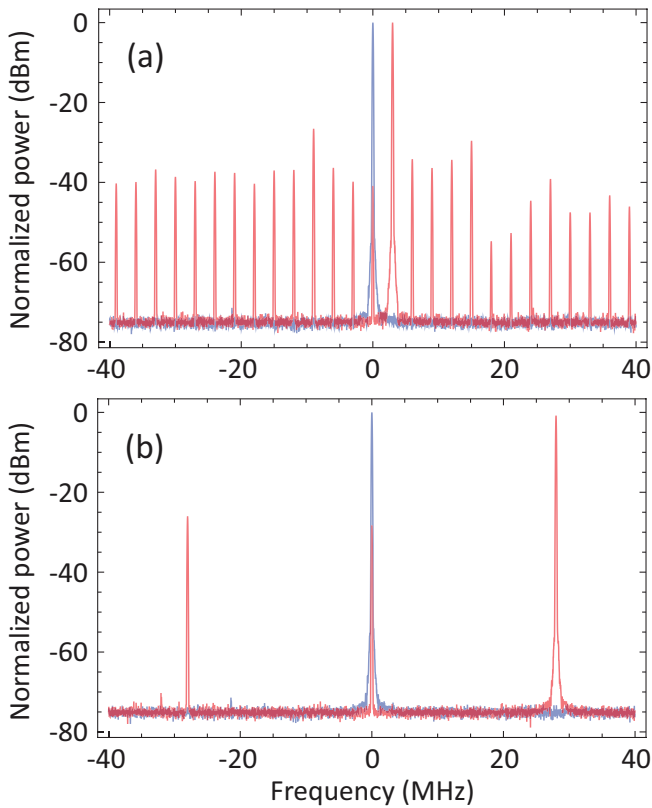}
\caption{Log plots of the reconstructed optical spectra of the serrodyne-modulated beam and the unmodulated carrier. The modulated spectrum (red curve) is normalized to the unmodulated carrier (blue curve) power. (a) $3\; \text{MHz}$ modulation. The efficiency is $98.5\%$, with the carrier suppressed to $-41\; \text{dB}$ below peak power and the largest sideband suppressed to $-26.6\; \text{dB}$. (b) $28\; \text{MHz}$ modulation with efficiency  $96.5\%$, carrier suppression of $-30\; \text{dB}$, and largest sideband suppression of $-26\; \text{dB}$.}
\label{fig:serrodyne_modulation}
\end{figure}

The frequency of each beam can be shifted by phase modulation of the $1560\; \text{nm}$ lasers via the EOMs.  For a carrier signal with amplitude $E_0$ and frequency $\omega_0$, the phase-modulated signal is given by 
\begin{equation} \label{phasemod}
E = E_{0}\exp\left[i\left(\omega_{0}t+h\;\phi(t)\right)\right]
\end{equation} 
where $h$ is the modulation depth and $\phi(t)$ is the modulation waveform.  
We employ serrodyne modulation \cite{johnson2010broadband} to suppress undesired sidebands. The idea of serrodyne modulation is to use 
\begin{equation} \label{serrodyne}
\phi(t) = \omega_m t \; \text{mod}\; (2\pi / h)
\end{equation}
as the modulation waveform, where $\omega_m$ is the desired frequency shift.  In the ideal case, serrodyne modulation transfers all the carrier power to a single sideband with frequency $\omega_0 + \omega_m$.  However, bandwidth limitations provide a lower bound on the fall time of the sawtooth waveform described by Eq.~\ref{serrodyne} and lead to additional sidebands in the phase-modulated spectrum.  The efficiency of serrodyne modulation thus depends on the bandwidth of the applied waveform.

In our apparatus, the sawtooth waveform is generated by an arbitrary waveform generator (AWG, Tektronix AWG5014) with an analog bandwidth of $230\; \text{MHz}$. The voltage output level of the AWG is sufficient to reach $\text{V}_{\pi}$ of the EOMs without the use of an external amplifier. Note that the voltage required to achieve the serrodyne condition for a frequency doubled system is half of what is required for a non-doubled system.
Fig. \ref{fig:serrodyne_modulation} shows a log plot of the unmodulated input and the modulated output spectra for $3\; \text{MHz}$ and $28\; \text{MHz}$ modulation. Negative-frequency modulation can be carried out by applying a negative-slope sawtooth from the negative output port of the AWG. For $3\; \text{MHz}$ modulation, the transfer efficiency is $98.5\%$ with the carrier suppressed by $-41\; \text{dB}$ and the largest sideband suppressed by $-26.6\; \text{dB}$. For $28\; \text{MHz}$ modulation, the efficiency is $96.5\%$, the carrier suppression is $-30\; \text{dB}$, and the largest sideband is suppressed by $-26\; \text{dB}$. In both cases, the amplitude of the sawtooth modulation was $2.29\; \text{V}_\text{pp}$.  The serrodyne modulation was tested at frequencies of up to 400 MHz, where the efficiency is $63.3\%$ at a modulation amplitude of $4.12\; \text{V}_\text{pp}$.

The modulation angular frequency retains its original value even after the modulated beam is frequency doubled \cite{boyd2019nonlinear}. This can be seen from the fact that the frequency-doubled field is proportional to the square of the input field. Squaring Eq.~\ref{phasemod}, we obtain
\begin{equation}
\label{eq:doubled_mod}
     E^2 = E_{0}^{2}\exp\left[i\left(2\omega_{0}t+2h\;\phi(t)\right)\right].
\end{equation}
Eq. \ref{eq:doubled_mod} indicates that when the modulated signal is frequency doubled, the carrier frequency and the modulation depth are doubled, but the modulation frequency $\omega_{m}$ remains unchanged.

We implement a phase lock to stabilize the relative phase of the two beam paths. The error signal consists of the beating signal obtained by interfering a small fraction of the light from each path after the high-power fibers.  This signal is amplified, demodulated, low-pass filtered, and used as the input of a PID controller (Liquid Instruments Moku:lab PID controller). The PID controller has an input bandwidth of $200\; \text{MHz}$ and an output bandwidth of $>300\; \text{MHz}$ ($3\; \text{dB}$ point). The error signal and output of the PID controller are connected to a SPDT switch to disable the feedback while the light is off.  The output of the PID controller is amplified and applied to the corresponding EOM.

The performance of the phase lock was measured for continuous wave (CW) operation and for pulses of duration $90\; \mu$s to $2\; \text{ms}$ at repetition rates of $100\; \text{Hz}$ to $1\; \text{kHz}$, the timescales relevant for driving sequential two-photon transitions. Each pulse is initially turned on at a lower power for $20\; \mu$s to allow the phase lock to acquire. In CW operation, we observe relative phase noise below $-91\; \text{dBc/Hz}$ at an offset of $10\; \text{kHz}$.

In summary, we have developed a 780 nm laser system with up to $40\; \text{W}$ in frequency components separated by $370\; \text{GHz}$.  These frequency components are efficiently overlapped on PPLN nonlinear crystals. Serrodyne modulation using a high-bandwidth sawtooth waveform allows us to perform highly efficient frequency shifts. This laser system is expected to improve the performance of current and future high-precision measurements with atom interferometry. Other experiments that rely on high power lasers with frequency tunability, such as trapped ion experiments \cite{blatt2008entangled}, can also benefit from this technique. The number of frequency components in each beam could be increased by including additional nonlinear crystals. 

We acknowledge funding from the Office of Naval Research DURIP and the Vannevar Bush Faculty Fellowship Program. We thank Agnetta Cleland and Megan Nantel for their assistance with this work.

\bibliographystyle{apsrev4-1}
\bibliography{reference}

\end{document}